Transition from high-entropy to conventional (TiZrNbCu)$_{1-x}$Co$_x$ metallic glasses

Ramir Ristić[a*], Ignacio A. Figueroa[b], Amra Salčinović Fetić[c], Krešo Zadro[d], Vesna Mikšić Trontl[e], Petar Pervan[e], Emil Babić[d**]

[a*]Department of Physics, University of Osijek, Trg Ljudevita Gaja 6, HR-31000 Osijek, Croatia

[b]Institute for Materials Research-UNAM, Ciudad Universitaria Coyoacan, C.P. 04510, Mexico D.F., Mexico

[c] University of Sarajevo, Faculty of Science, Zmaja od Bosne 35, 71000 Sarajevo, Bosnia and Herzegovina

[d]Department of Physics, Faculty of Science, Bijenička Cesta 32, HR-10002, Zagreb, Croatia

[e] Institute of Physics, Bijenička Cesta 46, P.O.Box 304, HR-10 001, Zagreb, Croatia

Corresponding authors:

[*]E-mail address: ramir.ristic@fizika.unios.hr. (R. Ristić)

[**]Email address: ebabic@phy.hr (E. Babić)

Abstract

A new amorphous alloy system (TiZrNbCu)$_{1-x}$Co$_x$ covering a broad composition range from the high-entropy (HEA) to Co rich alloys ($x≤0.43$) has been fabricated, characterized and investigated. A comprehensive study of the chemical compositions, homogeneity, thermal stability, electronic structure and magnetic and mechanical properties has been performed. All properties change their variations with $x$ within the HEA range. In particular, the average atomic volume deviates from the Vegard´s law for $x>0.2$, where also the average atomic packing fraction suddenly changes. The valence band structure, studied with ultraviolet photoemission spectroscopy, shows a split-band shape with 3d-states of Co approaching the Fermi level on increasing $x$. Due to onset of magnetic correlations magnetic susceptibility rapidly increases for $x>0.25$. Very high microhardness increases rapidly with $x$. The results are compared with those for similar binary and quinary metallic glasses and with those for Cantor type of crystalline alloys.

1.Introduction

New alloy design based on multi-principal element solid-solutions has been introduced at the beginning of this century. First applied to amorphous alloys [1,2] in an effort to discover new bulk metallic glasses (BMG) this design soon spread to crystalline alloys (so-called high-entropy alloys, HEA [3,4]) becoming a forefront of research in materials science [5,6]. HEAs and BMGs share some common features, such as elemental disorder and quite often phase metastability which strongly characterize all their properties. Since HEA design in contrast to that of conventional alloys (CA) explores the middle section of the phase diagrams of multicomponent alloys it makes a virtually unlimited number of new alloys available for research and possible application [5,6,7]. This unique opportunity to advance our knowledge and achieve the industrial application of compositional complex alloys (CCA) aroused huge research effort into design, fabrication and studies of HEAs which resulted so far in several hundreds of new alloys, thousands of research papers, dozens of the reviews of the literature (e.g.,[8-14]) and two books [15,16].

As result of all these efforts, large progress has been made in the knowledge of HEAs. Several technologically relevant alloys, such as those with excellent low- and high-temperature mechanical properties [5,13] have been discovered. At present research of HEAs is mostly oriented towards the development of new structural materials, thus focused on their mechanical properties and microstructures [5,6,10,12,13,15,16]. Accordingly, the majority of studied HEAs are based on the iron group of elements, followed to lesser extent by refractory ones.

At the same time, the research of their physical properties is despite of their potential as functional materials [5], still insufficient. The frequent lack of detailed insight into their electronic structures (ES), which, in metallic systems determine all intrinsic properties [17-34] including the mechanical ones [5,14,20] is possibly the main hindrance to the conceptual understanding of both crystalline (c-) and amorphous (a-) HEAs. Two other important issues in CCAs that have hardly been studied so far are the transition from HEA to CA with the same chemical make-up [17,18,27,29,35-38] and the disentanglement of the effects of topological and chemical disorder on their properties [17,24,39-44]. The study of transition from HEA to CA is important, both for understanding the formation of HEAs and for proper evaluation of their potential with respect to that of the corresponding CAs.

The alloys composed from early (TE) and late (TL) transition metals [17,24,39-41] are advantageous in studying all these issues since they possess a relatively simple split-band ES, strong interatomic bonding and large atomic size mismatch (which favour the elemental and topological ordering under suitable conditions [16,39,40]) and can be prepared in an amorphous phase [45] over a broad composition range which facilitates the study of the transition from HEA to CA in the same alloy system [17,18].

Accordingly, we performed recently comprehensive study of the effect of transition from HEA to CA concentration range on properties of $(TiZrNbCu)_{1-x}Ni_x$ and $(TiZrNbNi)_{1-x}Cu_x$ (for short, Ti-Zr-Nb-Cu-Ni) metallic glasses (MGs) [17,18,27,29]. The main result of this study is that a simple exchange of selected principal element Ni with Cu makes a drastic change on variations of properties with $x$, thus in the nature of the transition from a-HEA to Ni- or Cu-rich conventional MGs. In particular, the properties of $(TiZrNbNi)_{1-x}Cu_x$ MGs[18] showed, like binary TE-Cu MGs [46] and $(CrMnCoNi)_{1-x}Fe_x$ face centered cubic (fcc) solid solutions [37,38] an ideal solution behaviour associated with linear variations of properties with $x$, thus no real change on crossing from HEA to CA concentration range. In contrast, $(TiZrNbCu)_{1-x}Ni_x$ MGs [17,27,29], like $(CrMnFeCo)_{1-x}Ni_x$ fcc solid solutions [36,37], showed a strong change in properties on crossing from HEA to CA concentration range.

Here, we present results of the comprehensive study of properties of a new alloy system $(TiZrNbCu)_{1-x}Co_x$ MGs, performed over a broad composition range $x \leq 0.43$. This system is selected to elucidate the influence of the atomic number of TL constituent on the properties of quinary TE-TL MGs. We studied the thermal stability parameters, atomic structure, ES using ultraviolet photoemission spectroscopy, magnetic susceptibility and microhardness. To our knowledge, the study of microhardness is the first such study performed over the composition range extending from HEA to CA concentration range. The main result is that all these properties change their variation with Co content around $x=0.25$, thus deep within the HEA concentration range. Around this composition, thermal parameters change a slope of their variation with $x$, atomic volumes deviate from Vegard´s law behaviour and magnetic susceptibility starts to increase with Co-content. Thus, in quinary TE-TL MGs the TL content at which the properties change their variation increases with the increasing atomic number of TL element. Based on an inspection of the compositional variation of the criteria for glass forming ability (GFA) of our alloys we propose a new explanation for the discrepancy between the high thermal stability of a-HEAs and their modest GFA. The results are compared with those for similar binary TE-TL MGs [47-49], quinary Ti-Zr-Nb-Cu-Ni MGs [17,18,27] and with those for Cr-Mn-Fe-Co-Ni fcc solid solutions [36-38].

2. Experimental procedures

In preparing the $(TiZrNbCu)_{1-x}Co_x$ alloys we followed the same procedures used previously for the preparation of Ti-Zr-Nb-Ni-Cu alloy systems [17, 18, 27]. In particular, the ingots of seven alloys with $x=0, 0.1, 0.2, 0.25, 0.32, 0.43$ and $0.5$ were prepared from high purity components ($\geq 99.8$ at. %) by arc melting in high purity argon in the presence of a titanium getter. All ingots were flipped and remelted five times to ensure complete melting and good mixing of components. The samples in a form of ribbons with a thickness of about 25 micrometers of each alloy were fabricated by melt-spinning molten alloy on the surface of a copper roller rotating at the speed of 25 m/s in a pure helium atmosphere. Casting with controlled parameters resulted in ribbons with similar cross-sections and surface appearance, thus with the amorphous phases having a similar degree of quenched in disorder.

All as-cast ribbons were studied by X-ray diffraction (XRD) using a powder diffractometer with a Cu-Kα source [17, 18, 24, 27]. The XRD patterns showed that all samples, except for those with $x$=0 and 0.5 were amorphous. (In a separate study [50] the atomic structure of amorphous samples was also investigated using a synchrotron based high-energy X-ray diffraction (HEXRD) at the Diamond Light Source, Didcot, UK.) The ribbons which appeared in X-ray amorphous form were further studied by differential scanning calorimetry (DSC) and thermogravimetric analysis (TGA) using a Thermal Analysis DSC-TGA instrument. Thermal measurements were performed up to 1600 K with a ramp rate of 20 K/min. The values of thermal parameters were determined by using TA "Advantage" software. The regular calibration of DSC-TGA equipment [17,18,24] keeps the uncertainty in measured temperatures to within +/- 5 K. Fully amorphous as-cast ribbons were also investigated with scanning electron microscopy (SEM) using a JEOL ISM7600F microscope with energy dispersive spectroscopy (EDS) capability in order to determine their actual compositions and chemical homogeneity [17, 24]. Elemental mapping was performed on three different areas of each sample.

The valence band structure of all as-cast amorphous alloys was studied by ultraviolet photoemission spectroscopy (UPS), with a Scienta SES100 hemispherical electron analyzer attached to an ultra-high vacuum chamber with a base pressure below $10^{-9}$ mbar [18,29]. An unpolarized photon beam of energy 21.2 eV was obtained by a He-discharge ultraviolet photon source. The samples were cleaned by several cycles of sputtering with a 2 keV $Ar^+$ ions at room temperature to remove the oxygen and other contaminants from the surface. The overall energy resolution in experiments was about 25 meV.

The as-cast ribbons were also used for measurements of magnetic susceptibility and preliminary measurements of the low temperature specific heat (LTSH)[51]. The magnetic susceptibility of all alloys was measured with a Quantum Design magnetometer, MPMS5, in a magnetic field B up to 5.5 T and temperature range 5-300 K [17,18,21,22,27]. Since the magnetic susceptibility of all samples, except for that with $x$=0.5, showed a weak dependence on the temperature within the explored temperature range (as is usual in nonmagnetic alloys of TE and TL metals [17,18,21,22,46,52]) in the following analysis we will use the room temperature values. The preliminary measurements of LTSH on two samples with $x$=0.2 and 0.43 [51] were performed in the temperature range 1.8-300 K using a Physical Property Measurement System (PPMS) Model 6000 from Quantum Design, as previously described [17,18,24,27].

The microhardness of all as-cast amorphous samples has been measured at room temperature. These measurements were performed using DHV-1000Z Micro Vickers Hardness Tester device equipped with pyramidal indenter with square base, having an angle $137^0$. Ten indentations were made on both sides of each sample. The loading time was 15 s and the load was 0.981 N. The standard deviations were about 5% of the mean values.

3. Results and discussion

3.1 Thermophysical parameters and elemental mapping

A large effort has been devoted to the prediction of the phase formation in HEAs [5-17]. Several semiempirical criteria for the formation of different phases, single phase solid solution (SS), intermetallic compounds (IM) and their mixture with SS and /or amorphous phase, a-HEA, have been proposed. These criteria are mainly based on thermophysical parameters such as the mixing or formation enthalpy [53], $\Delta H_{mix}$, the ideal configurational entropy, $\Delta S_{conf}$, the atomic size mismatch, $\delta$, valence electron concentration, *VEC*, etc. (see e.g., [11,12]). Despite of their limitations and some erroneous predictions (e.g.,[11,16,18,40]) these criteria are useful for a quick comparison of different HEA systems [18], while the variation of thermophysical parameters with composition within a given alloy system can provide an insight into the evolution of this system [17,18].

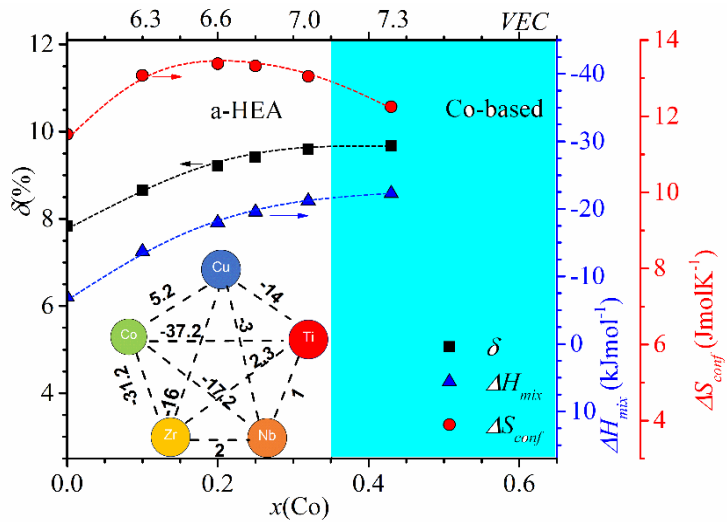

Fig. 1. Thermophysical parameters of (TiZrNbCu)$_{1-x}$Co$_x$ alloys vs. *x* and valence electron count *VEC* (upper abscissa). The inset: $\Delta H_{mix}$ between constituent elements [53].

In Fig. 1 we show the variations of the thermophysical parameters of all studied (TiZrNbCu)$_{1-x}$Co$_x$ alloys with Co content. In the calculation of these parameters we used standard expressions (see, e.g., [11, 17, 18]) and the input parameters for $\Delta H_{mix}$ and $\delta$ in Fig.1 and inset were taken from Ref. 53 and 54, respectively. In Fig. 1 the concentration range of HEAs, $x \leq 0.35$, is distinguished from that of Co-rich alloys by a different color. The values of *VEC*, which are proportional to *x* are shown on the upper abscissa. The range of values of $\Delta H_{mix}$ (from -22 kJmol$^{-1}$ to -6.8 kJmol$^{-1}$) and $\delta$ (from 7.8 % to 9.7%) place our alloys in a standard $\Delta H_{mix}$-$\delta$ plot within the region occupied with IM (*x*=0) and a-HEAs (higher x) [11], which is consistent with our experimental findings. As can be seen from the inset, small values of $\Delta H_{mix}$ are the consequence of strong interatomic interactions between TE and TL atoms [53] and large $\delta$ is similarly due to the large difference in size between TE and TL atoms. Since small $\Delta H_{mix}$ and large $\delta$ are general features of TE-TL alloys this facilitates the comparison of the results for our quinary MGs with previous results for similar binary ones [47-49].

The variations of $\Delta H_{mix}$ and $\delta$ with composition in our alloys shown in Fig. 1 are quite similar to these in (TiZrNbCu)$_{1-x}$Ni$_x$ alloys [17]: they vary rapidly at lower values of $x$ and tend to reach shallow maxima in a Co rich concentration range around $x$=0.5. However, $\Delta H_{mix}$ of Co-rich alloys is somewhat larger than that in corresponding Ni-rich alloys [17] which is a consequence of both, higher values of $\Delta H_{mix}$ between Co and the employed TE atoms and higher $\Delta H_{mix}$ between Co and Cu, than these of Ni with the same TE and Cu atoms (inset to Fig. 1). (Higher $\Delta H_{mix}$ between given TEs and Co than that for Ni is consistent with a trend in transition metals that $\Delta H_{mix}$ decreases on the increasing distance between two elements in the periodic table [53].) Rather strong interactions between Co and TE atoms, the demixing tendency between Co and Cu and a high melting point of Nb can all affect the distribution of constituents in our alloys.

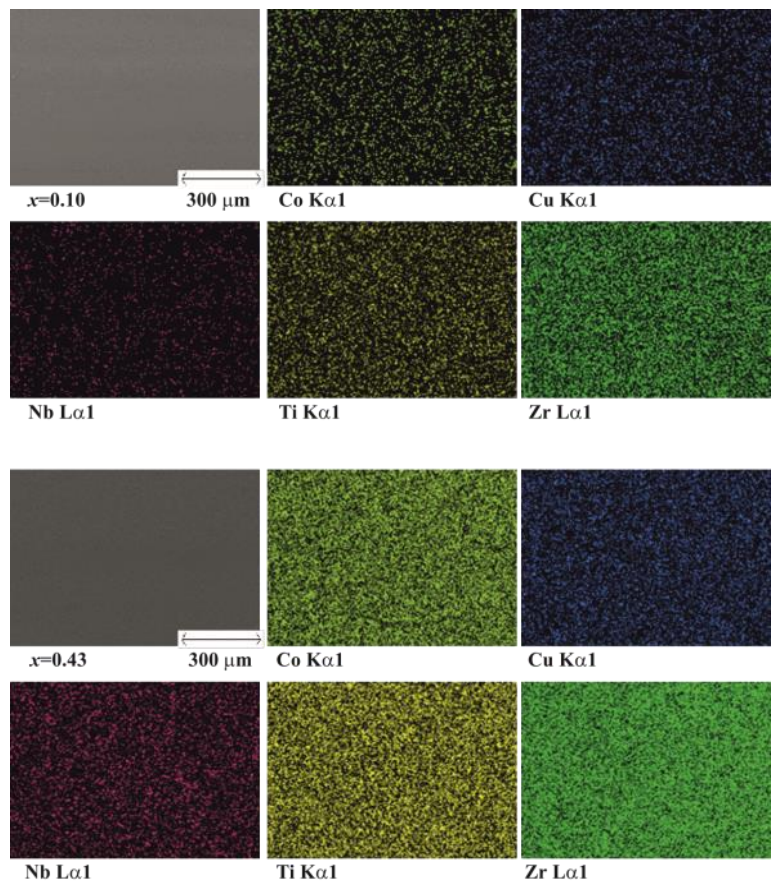

Fig. 2. SEM/EDS images for as-cast (TiZrNbCu)$_{1-x}$Co$_x$ alloys with $x$=0.1 and 0.43.

Careful characterization of compositional homogeneity of HEAs is particularly important since an uneven distribution of constituents can occur even in HEAs that are SS according to their XRD pattern [8,55]. Accordingly, we have performed EDS mapping of the distribution of constituent elements in all our as-cast MGs. Fig. 2 shows the SEM images and corresponding elemental maps of the constituent elements in our alloys with $x$=0.1 and 0.43, respectively, thus covering both HEA and Co rich concentration range. As described in our previous papers [17,18,24], the elemental mapping was performed on three different areas of each alloy in order to access the eventual inhomogeneity in the distribution of the constituents, through the variation of the composition in different areas of the same sample and also to obtain some information about the possible size and shape of such inhomogeneity (e.g.,[39]). As illustrated in Fig. 2 for MGs with $x$=0.1 and 0.43, the distributions of all constituent elements were random down to micrometer scale in all our alloys. We did not observe any clear correlation or anticorrelation between distributions of different elements. Although the elemental mapping in Fig. 2 cannot exclude some compositional fluctuations on a nanometric scale, such fluctuations, even if present, are not likely to have any larger effect on the macroscopic (bulk) properties studied by us. Further, in all samples, the composition calculated from EDS at different locations was the same within about 1 at. % which is an indication of their macroscopic homogeneity. Since the average concentrations of our alloys obtained from EDS were within about 1 at. % of the corresponding nominal ones, we will continue to use the nominal compositions in our further analyses.

3.2 Thermal parameters

Thermal parameters are particularly important since they determine the useful temperature range in all alloys (e.g.,[7,12,57]). Further, these parameters are related to the strength of interatomic bonding in an alloy. It is therefore surprising that the thermal analysis of HEAs is often ignored and thermal parameters are estimated by using the rule of mixtures, ROM [7] which is known to provide erroneous values of these parameters in both c-HEAs (e.g.,[57]) and a-HEAS [17,18,24,28]. We note however that for some alloy systems, such as these based on refractory elements, the temperature range of commercial DSCs (usually T≤ 1400 $^0$C) may not be sufficient for their complete thermal analysis.

Fig. 3 shows the results of the thermal analysis on all (TiZrNbCu)$_{1-x}$Co$_x$ alloys exhibiting an amorphous XRD pattern. The thermal analysis confirmed the XRD results in that all alloys with 0.1 ≤ $x$≤0.43 were amorphous from which we could extract the values of their thermal parameters: the glass transition ($T_g$), crystallization ($T_x$), melting ($T_m$) and liquidus ($T_l$) temperatures. The DSC/DTA traces in Fig. 3 are quite similar to those observed recently in Ti-Zr-Nb-Cu-Ni MGs [18,27]. All alloys show rather complex crystallization patterns reflected in three or more exothermic maxima spread over a broad temperature range. This is consistent with a strong bonding tendency between the alloying elements inferred from thermophysical parameters in Fig. 1. Rather complex crystallization processes will make future studies of the evolution of crystallization products with composition and temperature much more complicated than these for corresponding binary MGs [47]. For the sake of simplicity, here we analyze only the temperatures of first crystallization event , which determines the stability of the amorphous phase.

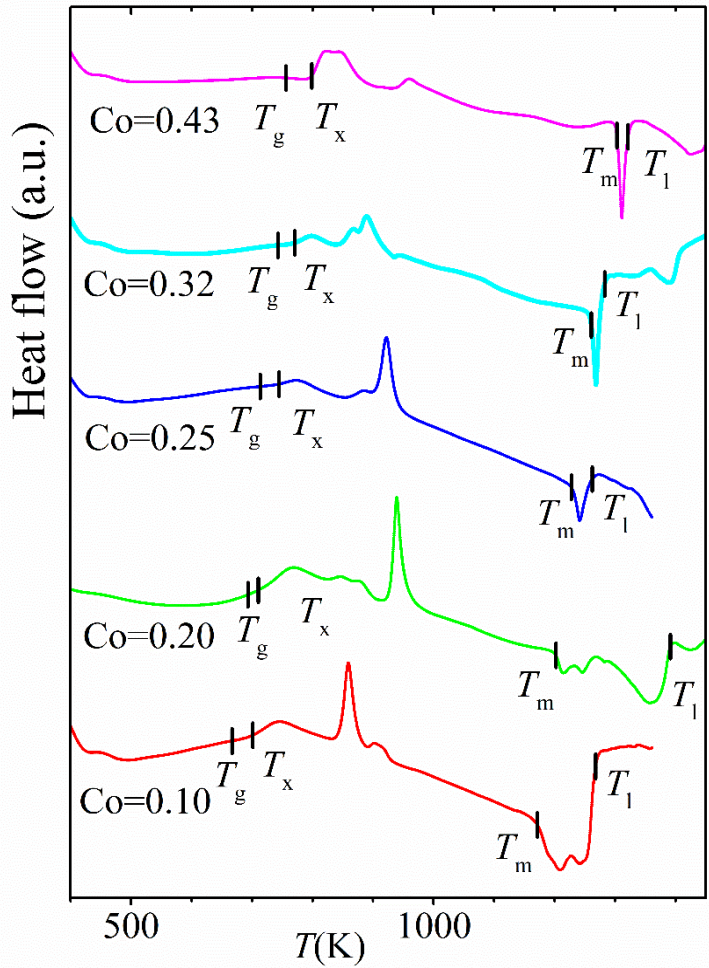

Fig. 3. DSC/DTA traces of $(TiZrNbCu)_{1-x}Co_x$ alloys. Note complex crystallization patterns.

The variations of $T_l$, $T_m$, $T_x$ and $T_g$ with Co content $x$ are shown in Fig. 4. These variations delineate a nonequilibrium phase diagram [18] of our alloy system in which different colours denote different phases. Above $T_l$ the alloy is in a liquid state, and in the temperature interval between $T_l$ and $T_m$ a coexistence of solid phases and liquid is established. Between $T_m$ and $T_x$ ($T_x$ denoting the onset of the first crystallization event) alloys are in a crystalline state, whereas between $T_x$ and $T_g$ the system is a supercooled liquid. Below $T_g$ alloys enter into the amorphous phase. We note that all thermal parameters, associated with the thermal stability of different phases and interatomic bonding, increase with increasing Co content. This is the usual behaviour of $T_x$ and $T_g$ in binary and ternary TE-TL MGs [47,58-61,67] which was recently also observed in quinary Ti-Zr-Nb-Cu-Ni MGs [18,27]. Such behaviours of $T_x$ and $T_g$ are usually accompanied by a simultaneous enhancement of the mechanical properties and the Debye temperatures with increasing TL content, which all support the increase of the strength of interatomic bonding [17,27,46].

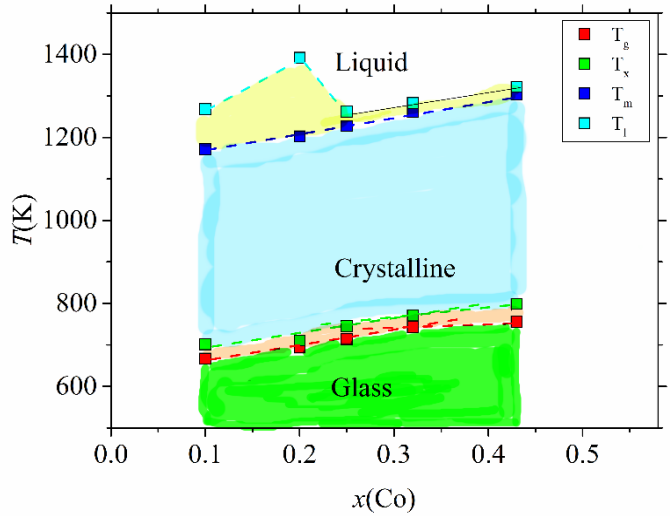

Fig. 4. Thermal parameters of $(TiZrNbCu)_{1-x}Co_x$ alloys.

However, the actual variations of thermal parameters with x in our alloys (Fig. 4) are somewhat different from those in binary [46] and quinary [18] TE-Cu alloys. In TE-Cu MGs quasi-linear variations of thermal parameters and other thermodynamic properties with Cu content indicate an ideal solution behaviour, thus the smooth transition from HEAs to Cu-rich conventional alloys in $(TiZrNbCu)_{1-x}Cu_x$ MGs [18]. However, in our alloys $T_x$ and $T_g$ seem to increase a little more rapidly with $x$ for $x \leq 0.25$ than at higher x (Fig. 4). This behaviour is similar to that observed in $(TiZrNbCu)_{1-x}Ni_x$ MGs where the increase of $T_l$ and $T_x$ in the HEA concentration range, $x \leq 0.35$, slowed down in Ni-rich alloys [27]. (In this system all studied properties, including the atomic short-range order (SRO) [17,27] and the electronic transport properties [29] changed their variations on the transition from HEA to Ni-rich conventional alloys.) However, $T_x$ of binary Zr-Co MGs [47] increased approximately linearly with Co content between 20 and 50 at % Co, thus a small change of a slope of variation of $T_x$ with $x$ around $x=0.25$ in Fig. 4 is specific to our quinary MGs.

We note a strong enhancement of $T_l$ in our equiatomic alloy (Fig. 4) which has not been observed in corresponding Ti-Zr-Nb-Cu-Ni alloys [18, 27]. This enhancement seems related to a very high and narrow last exothermic maximum of this alloy (Fig. 3), which probably reflects the formation of a particularly stable intermetallic compound. Simultaneously, $T_m$ increases approximately linearly with $x$ (Fig. 4), which probably indicates that the least thermally stable crystalline phase is of the same type in all our alloys and that its stability increases somewhat with increasing Co content. Like in Ti-Zr-Nb-Cu-Ni MGs [17,18,27] and crystalline $(CrMnFeCo)_{1-x}Ni_x$ alloys [36,57] the values of $T_m$ of our alloys calculated by using ROM [7] are at variance with the experimental ones. ROM predicts a linear decrease of $T_m$ from 2041 K for $x$=0 to 1924 K for $x$=0.43, whereas the experimental values increase from 1123 K to 1303.5 K over the same concentration range. Thus, ROM predicts both, too large values of $T_m$ and their wrong variation with $x$. The observed strong deviation of experimental values of $T_m$ from those calculated by using ROM is probably associated with a strong bonding tendency between alloying elements in our alloys (Fig.1). Therefore, it is possible that in some alloys composed from very similar elements with a weak mutual interatomic interactions ROM provides reasonable approximation for $T_m$.

By using the experimental values we can compare the contributions to the free energy from $\Delta H_{mix}$ and $\Delta S_{conf} T_m$ (Fig,. 1). As is usual in TE-TL alloys [17,18,24,27,39,40,46], $\Delta H_{mix}$ outweighs $\Delta S_{conf} T_m$ due to strong interatomic bonding in all our alloys containing Co ($x \geq 0.1$). Since our as-cast alloy TiZrNbCu in which $\Delta S_{conf} T_m$ outweighs considerably $\Delta H_{mix}$ (Fig. 1) was multiphase (IM) [17,24,27] it seems that in our alloys configurational entropy has limited influence on the formation of either SS or an amorphous phase.

The ratio $\Omega = \Delta S_{conf} T_m / \Delta H_{mix}$ where $T_m$ is calculated by using ROM is a commonly accepted criterion for the phase formation of HEAs [11,65]. Since the values of $T_m$ of our alloys calculated by using ROM are much larger than the experimental ones the corresponding values of $\Omega$ put all our alloys above the region occupied with amorphous alloys in the $\Omega$ vs. $\delta$ plot. (As already noted in section 3.1 our alloys are correctly placed in a $\Delta H_{mix}$ vs. $\delta$ plot [11].) This indicates that some erroneous predictions of phase by $\Omega$ criterion in both a- and c-HEAs (e.g.,[66]) may arise from the use of calculated instead of the observed value of $T_m$ in the definition of this criterion [65]. This emphasizes again the importance of the measurements of the thermal parameters of HEAs and other compositional complex alloys.

Thermal parameters are extensively used for the construction of the criteria for the glass forming ability, GFA of alloys [17,18,21,22,27,62-64]. The simplest criterion [63] which relates GFA to the reduced glass transition temperature:

$T_{rg} = T_g / T_l$        (1)

predicts good GFA for $T_{rg} \geq 2/3$. This criterion works quite well in binary and ternary TE-TL alloys [21,22,59-63], but seems to be less effective in the multicomponent bulk metallic glasses (BMG) and a-HEAs [18,24,27,62, 64]. $T_{rg} \geq 2/3$ requires low $T_l$ as is the case in deep eutectics and since $T_g$ varies slowly and approximately linearly with the composition (Fig. 4) the variation of $T_{rg}$ is determined by that of $T_l$ (e.g.,[67]). As shown in Fig. 5 this is also the case in (TiZrNbCu)$_{1-x}$Co$_x$ alloys since $T_{rg}$ reaches the lowest value at $x$=0.2, where $T_l$ is at its maximum (Fig. 4). This minimum of $T_{rg}$ is superposed on a convex variation of $T_{rg}$ with $x$ showing a shallow maximum around $x$=0.32 (Fig. 5). The variation of $T_{rg}$ with $x$ in Fig. 5 is similar to those in Ti-Zr-Nb-Cu-Ni MGs [18,28] and in these systems it was shown to be erroneous since the experiment showed the best GFA for equiatomiic TiZrNbCuNi alloy ($x$=0.2) having the lowest value of $T_{rg}$ [18]. Based on this finding [18] and on the fact that in all three quinary alloy systems the formation of glass becomes increasingly more difficult on approaching both, the lowest and the highest contents of TL ($x$) we expect that $T_{rg}$ criterion is unlikely to work in present alloys. However, modest values of $T_{rg}$≤0.58 in Fig. 5 are consistent with the observed modest overall GFA of our alloys.

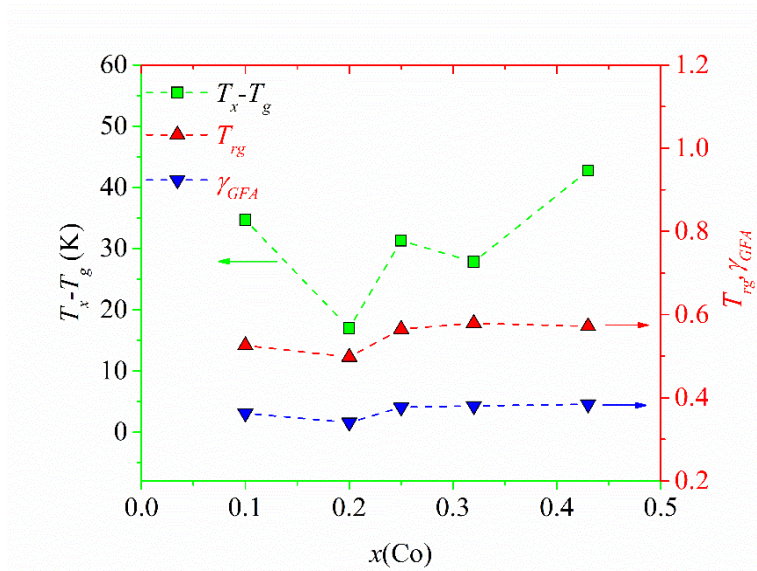

Fig. 5. Glass forming ability parameters (see text) of (TiZrNbCu)$_{1-x}$Co$_x$ alloys.

The erroneous prediction of GFA in our and other a-HEAs when using the $T_{rg}$ criterion may arise from their complex crystallization behaviour, involving several consecutive crystallization events (Fig.3). Accordingly, in all these systems different crystallization events (phases) determine the stability of the glassy state (limited by the first crystallization event, $T_x$) and $T_l$ (associated with melting of the most stable crystalline compound, probably formed at the later stage of crystallization, Fig. 3). This is different from the situation encountered in the majority of binary and ternary MGs [21,22,47,59-64,67] where there is a single exothermic maximum, thus the same crystalline phase(s) control both, the stability of glass against crystallization and the values of $T_m$ and $T_l$. Thus, complex crystallization patterns can provide an additional explanation of the discrepancy between the rather high stability and modest GFA of a-HEAs [24,56].

$T_{rg}$ ignores the stability of glass in respect to crystallization (Eq. (1)), which can be described with the width of the supercooled liquid region, $\Delta T_x = T_x - T_g$. Large $\Delta T_x$ is important for the application of MGs [62] and has also been proposed as one of the criteria for good GFA [45], but the correlation between the two was found to be statistically quite weak [62,64], More recently, $\Delta T_x$ was found to correlate with GFA of some a-HEAs better than $T_{rg}$ [18,56]. However, as seen in Fig. 5, $\Delta T_x$ does not seem to describe properly the variation of GFA with $x$ in our alloys since it shows a minimum at $x$=0.2 and achieves the largest values for $x$=0.1 and 0.43 (which are at the boundaries of the glass forming range in this system, thus unlikely to posses good GFA). The values of $\Delta T_x$ in Fig. 5 are on average about two times smaller than these of (TiZrNbNi)$_{1-x}$Cu$_x$ MGs [18], which indicates reduced stability of the glassy phase and probably poor overall GFA of our alloys.

In Fig. 5 we also show the variation of $\gamma$ criterion for GFA [62,64]:

$\gamma = T_x/(T_l+T_g)$ (2)

This criterion combines both, the kinetic effects in the denominator and the stability of glass in the numerator and correlates better with GFA in BMGs than the $T_{rg}$ and $\Delta T_x$ criteria [64]. In our alloys, $\gamma$ slowly increases with $x$, but shows a shallow minimum at $x$=0.2 (Fig. 5), similar to that of $T_{rg}$. Such variation of $\gamma$ is unlikely to provide a correct description of GFA in our system since we were unable to vitrify alloys with $x$>0.43. The failure of both, the $\gamma$ and $T_{rg}$ criteria for prediction of GFA in all our quinary MGs ([18,27] and Fig. 5) is probably caused by complex crystallization patterns of these alloys, as was illustrated in the case of $T_{rg}$ for present alloys. However, a modest maximum value of $\gamma$ in Fig. 5. $\gamma \leq 0.38$ is consistent with a poor GFA [62,64] of our alloys.

3.3 Atomic structure

XRD patterns of the melt-spun (TiZrNbCu)$_{1-x}$Co$_x$ ribbons spanning the whole concentration range, $x \leq 0.43$ are shown in Fig. 6. All alloys show a broad first maximum characteristic of amorphous alloys around $2\Theta=40^0$. These maxima shift to larger values of $2\Theta$ with increasing $x$ as could be expected due to the small size of Co atom [54]. The XRD patterns of metallic glasses in addition to showing their glassy nature can also provide an insight into the average interatomic distances, local atomic arrangements, average atomic volumes and the average atomic packing fractions, APF (e.g., [24,27,46,68]). Recently, we used the corresponding procedures to determine the change in the parameters associated with the local atomic arrangements accompanying the transition from HEA to Ni-rich or Cu-rich concentration range in (TiZrNbCu)$_{1-x}$Ni$_x$ and (TiZrNbNi)$_{1-x}$Cu$_x$ MGs [17,18,27], respectively. Here we use the same procedures to gain an insight into the evolution of local atomic structure with Co content in our alloys.

From the modulus of the scattering vector $k_p$, corresponding to the first maximum in the XRD pattern (Fig. 5), $k_p=4\pi\sin\Theta/\lambda$ ($\Theta$ is the Bragg angle and $\lambda$ is the wavelength of the employed X-ray radiation), one can calculate the average nearest neighbour distance [68]:

$d=7.73/k_p$ (3)

The values of $d$ of our alloys decreased with increasing $x$ and the slope of this decrease became a little larger for $x \geq 0.25$. The search for a local atomic arrangement and the calculation of average atomic volume from $d$ require the selection of a suitable atomic structure [17,18,24,27], such as the body-centred cubic (bcc), or face centred cubic (fcc) structure. For similar $(TiZrNbCu)_{1-x}Ni_x$ and $(TiZrNbNi)_{1-x}Cu_x$ MGs some facts, such as the bcc crystalline phase which crystalizes first upon annealing of alloys with lower Ni contents [39], as well as a good agreement between the calculated mass density obtained by assuming a bcc like atomic structure and the measured one [24] provided strong support for a bcc like local atomic structure. In addition to these facts, we note that a large difference between the size of TE atoms and TL ones [54] also makes the formation of a bcc like local atomic arrangement in TE-TL alloys more likely than the fcc one [11,39,40,55]. Since the thermophysical parameters (Fig. 1), thermal parameters (Figs. 3 and 4) and the electronic structure of present alloys are quite similar to those of $(TiZrNbCu)_{1-x}Ni_x$ and $(TiZrNbNi)_{1-x}Cu_x$ MGs it seems plausible to assume a bcc like local atomic arrangements also for our alloys. Accordingly, we have calculated the corresponding average bcc lattice parameters, $a_{bcc}=2d/3^{0.5}$ and average bcc atomic volumes, $V_{bcc}=a_{bcc}^3/2$ for all our alloys. The representation of the atomic parameters in terms of the atomic volume, $V_a$ seems advantageous in respect to that in terms of $d$, or $a$, since the value of $V_a$ calculated from the XRD pattern can be simply checked against that obtained independently from the mass-density, $D$ [69]:

$$V_a=1.66\ M/D \qquad (4)$$

where M is the composition averaged molar mass of the alloy.

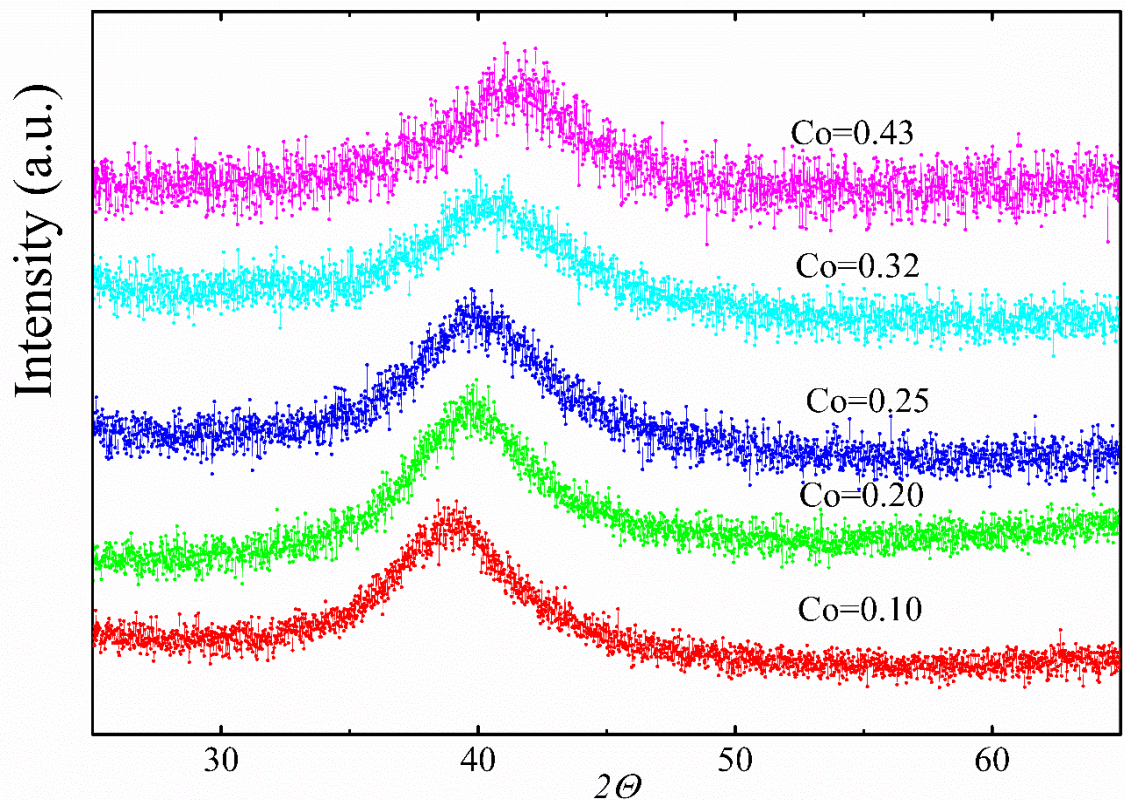

Fig. 6. XRD patterns of as-cast $(TiZrNbCu)_{1-x}Co_x$ alloys showing their amorphous structure.

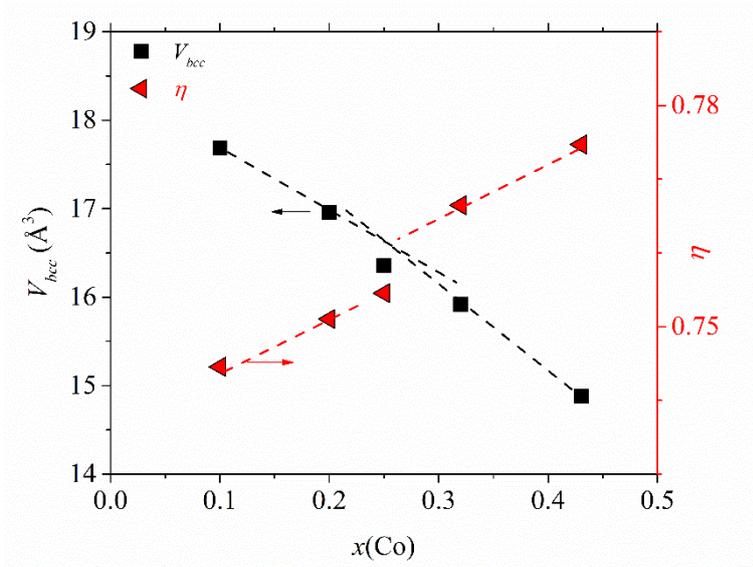

Fig. 7. Average atomic volumes $V_{bcc}$, and local atomic packing fractions $\eta$ of as-cast $(TiZrNbCu)_{1-x}Co_x$ alloys vs. $x$. Note change in variations around $x=0.25$.

As seen in Fig. 7 the variation of $V_{bcc}$ with $x$, like that of $d$, shows a small change of slope for $x \geq 0.25$. The values of $V_{bcc}$ in Fig. 7 are a little larger but close to corresponding values of $V_a$ calculated by assuming the validity of the Vegard's law. Moreover, a linear extrapolation of our data for $x \geq 0.25$ to $x=1$ yields $V_a$ around 11 Å$^3$ which is close to that of Co atom [54,69]. The linear variations of $V_a$ resembling those calculated by assuming the validity of the Vegard's law were previously observed over an extended range of concentration in $(TiZrNbCu)_{1-x}Ni_x$ and $(TiZrNbNi)_{1-x}Cu_x$ MGs [17,18,24,27] and in a single-phase crystalline fcc alloys derived from the Cantor alloy [35-38]. However, in $(TiZrNbCu)_{1-x}Ni_x$ MGs the Vegard's law like variation of $V_{bcc}$ was limited to the HEA concentration range, $x \leq 0.35$ [17], whereas in $(TiZrNbNi)_{1-x}Cu_x$ ones such variation persisted throughout the explored concentration range, $x \leq 0.52$ indicating an ideal solution behaviour [18]. In crystalline fcc $(CrMnFeCo)_{1-x}Ni_x$ alloys [35] $V_a$ obeyed the Vegard's law only in Ni-rich alloys, $x \geq 0.5$, whereas in fcc $(CrMnCoNi)_{1-x}Fe_x$ alloys $V_a$ obeyed the same law throughout the concentration range of stability of a single fcc phase, $x \leq 0.5$ [37,38]. Thus, the variation of $V_a$ in our alloys (Fig. 6) is qualitatively similar to that of $(CrMnFeCo)_{1-x}Ni_x$ alloys [35,36]. We note that in all alloys in which the variation of $V_a$ with $x$ showed some change, this change was accompanied by the corresponding change in some other properties [17,27,29,35-38]. In present alloys, like in similar $(TiZrNbCu)_{1-x}Ni_x$ MGs [27], change in variation of $V_{bcc}$ is accompanied by a small change in variation of thermal parameters, $T_g$ and $T_x$ (Fig. 4).

The variation of a mass-density of our alloys calculated by using eq. (4) with $V_a = V_{bcc}$ has also shown a small change in the slope for $x \geq 0.25$. Further, we use our values of $V_a$ to calculate the average local APFs [17,18,46] which depend on the local atomic arrangements. We have calculated APFs, $\eta_a$ of our MGs from the expression [18]:

$$\eta_a = \sum \eta_k^0 x_k V_k^0 / V_a \quad (5)$$

where $\eta_k^0$ is the APF of k-th component of the alloy in its crystalline phase and $x_k$ and $V_k^0$ are its molar fraction and atomic volume in the crystalline phase, respectively. The values of APFs of our alloys, obtained by using $V_a=V_{bcc}$, are shown in Fig.7. Our APFs are, like these in $(TiZrNbCu)_{1-x}Ni_x$ and $(TiZrNbNi)_{1-x}Cu_x$ MGs [17,18] quite high, $\eta_a$= 0.74 to 0.77 indicating rather efficient local atomic packing. However, in contrast to nearly constant APFs in $(TiZNbNi)_{1-x}Cu_x$ MGs [18], APFs of our MGs show a sudden increase for $x>0.2$ ( Fig. 7) indicating some change in the local atomic arrangements. Accordingly, parameters associated with the atomic structure of our alloys seem to indicate a change in the local atomic arrangements occurring around $x=0.25$, thus within the HEA concentration range.

However, the employed procedure for the determination of average $d$ from XRD pattern (eq. (3)), although providing quite a good description of the atomic parameters in several of TE-TL MGs (e.g.,[17,18,46,68,69]) lacks firm theoretical foundations [70]. Since our data for $d$, obtained from eq. (3) form the basis for all results displayed in Fig. 7, it is useful to corroborate these results by the data from additional measurements which can provide more direct insight into the atomic SRO in MGs [17,18]. Very recently a detailed high energy X-ray diffraction study of our MGs has been performed [50]. In this study the values of $d$ obtained from the radial distribution functions, R($r$) showed a faster decrease on increasing $x$ for $x \geq 0.25$, thus confirm our results for $d$. Moreover, the average coordination number of the first coordination shell, $N$ determined from R ($r$) showed an increase for $x \geq 0.25$, which coincides with a sudden increase of our APFs in Fig. 7. We note that a similar increase of $N$ was observed previously in $(TiZrNbCu)_{1-x}Ni_x$ MGs, but for $x>0.35$, thus in the Ni-rich concentration range. The change in atomic SRO of these alloys for $x>0.35$ was accompanied by the change in variations of all studied properties of these MGs. Thus, we may expect a similar change in the physical properties of our alloys for $x \geq 0.25$.

3.4 Electronic structure and physical properties

It is well known that ES determines all intrinsic properties of materials [71]. Thus, the knowledge of ES is necessary both for understanding the properties of materials and for the design of new materials with desired characteristics. Indeed, ES was found to control the atomic structure and properties even in the dilute alloys [72]. As noted earlier [46-52,58-61,69] the relationship between ES and properties is particularly simple in the TE-TL MGs. The early UPS and XPS studies of these alloys [73-76] revealed a split band structure of their valence bands, VB with the d-subbands of TL elements having full or nearly full d-shell positioned well below the Fermi level, $E_F$. Therefore, as long as the subband of TL remains well below $E_F$, the effect of alloying with TL can be regarded as a dilution of amorphous TE which explains approximately linear variations of most properties of these MGs with TL content [46-52,58-61,69,77]. Further, it was shown that the split band shape of VB of TE-TL alloys also applies to crystalline alloys [76] and is rather insensitive to the number of the alloying components [17,18,29,74]. In the alloys of early and late transition metals, the shift from the Fermi level of d-band in 3d transition metals increases with the increasing atomic number of elements, i.e. the binding energy $E_B$ increases as we go from Mn to Cu [74]. Besides, a small decrease of $E_B$ is observed for a given TL when its relative content in the alloy is increased. Similar behaviour of $E_B$ was recently observed also in Ti-Zr-Nb-Cu-Ni MGs [18,29].

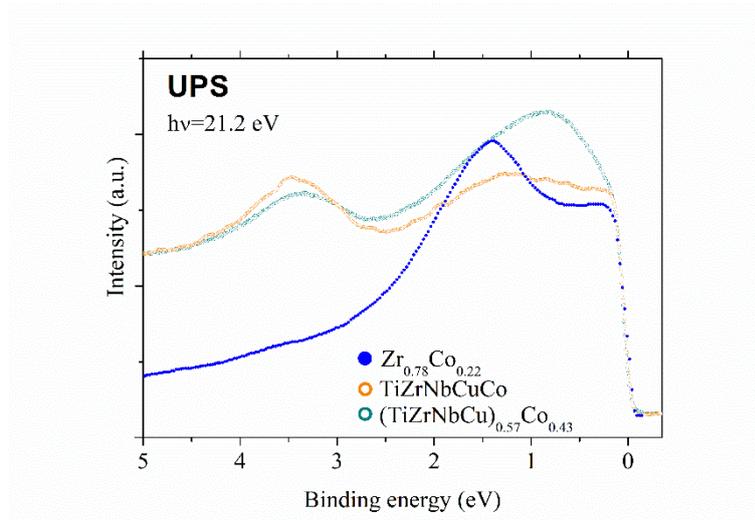

Fig. 8. Ultra-violet photoemission spectra of selected (TiZrNbCu)$_{1-x}$Co$_x$ and Zr$_{0.78}$Co$_{0.22}$ [73] amorphous alloys.

The results of UPS for our alloys with $x$=0.2 and 0.43, thus covering both HEA and Co-rich concentration range are shown in Fig. 8. For the sake of comparison the UPS data for binary Zr$_{78}$Co$_{22}$ MG [73] are also shown in Fig. 8. Taking into account the effect of photon energy-dependent photoemission cross-section [75] these spectra reflect the variation of the electronic density of states, DOS within the VB. Further, due to the generally low contribution of s,p-bands to the photoemission intensity these spectra largely reflect the DOS of d-electrons [17]. The spectrum for Zr$_{78}$Co$_{22}$ MG shows a peak around $E_B$=1.4 eV corresponding to the 3d states of Co and a small hump close to $E_F$ associated with 4d-states of Zr [73]. The corresponding spectrum for our alloy with a similar Co content, $x$=0.2, shows a peak centered around $E_B$=3.5 eV associated with the 3d states of Cu [17,18,29] and a hump around $E_B$=1.3 eV probably related to the 3d-states of Co. A poorly resolved maximum of 3d-states of Co in our alloy is probably due to the nearby maximum of 4d-states of Nb [17] as well as due to generally very large photoemission cross-sections of TEs at low photon energies [75]. Accordingly, in order to obtain more accurate information about the position and shape of the contribution of 3d states of Co in the DOS of this alloy the XPS measurements using higher photon energy are required.

However, on increasing Co fraction to $x$=0.43 the peak corresponding to 3d states of Co strongly increases and shifts towards $E_F$, $E_B$=0.8 eV (Fig. 8). We note that in similar Ti-Zr-Nb-Cu-Ni MGs the shift of $E_B$ of 3d-states of Cu and Ni for the same $x$=0.43 was much smaller than that in Fig. 8 [18,29]. Accordingly, we expect that a band crossing similar to that observed in Zr-Co MGs [48,49] will also occur in our alloys at a higher concentration of Co. Assuming that the expression used to estimate the crossover concentration at which the d- states of TL start to dominate $N(E_F)$ in Zr-TL MGs [48] applies also to our quinary MGs we find the crossover concentration $x$=0.68, which is considerably higher than that in Zr-Co MGs (around $x$= 0.5 [48,49]). The trend to the band crossing clearly visible in the UPS spectrum of the alloy with $x$=0.43 has also been confirmed in the results of preliminary LTSH measurements as well as those of magnetic susceptibility, described below. In particular, the dressed density of states, $N(E_F)$ increased from 1.7 (at. eV)$^{-1}$ to 2.0 (at. eV)$^{-1}$ from $x$=0.2 to $x$=0.43 [51]. Since the electron-phonon enhancement of $N(E_F)$ is likely to decrease on increasing Co content, the relative increase of the bare DOS at $E_F$ is probably even larger.

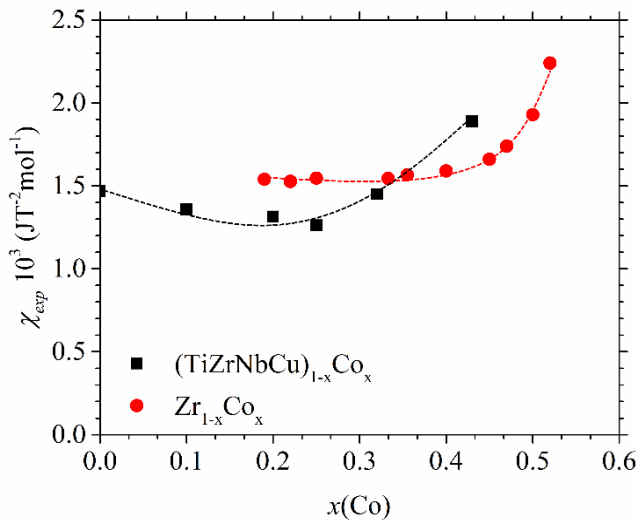

Fig. 9. Room temperature magnetic susceptibilities of (TiZrNbCu)$_{1-x}$Co$_x$ and Zr$_{1-x}$Co$_x$ [58,79] metallic glasses vs. $x$.

The variation of the magnetic susceptibility at room temperature, $\chi_{exp}$ with Co content for all our alloys is shown in Fig. 9. For the sake of comparison the corresponding data for Zr$_{1-x}$Co$_x$ MGs [58,69,78,79,80] are also shown in the same figure. Accidentally, the data for both alloy systems extrapolate for x=0 to the same value around 1.6 10$^{-3}$ JT$^2$mol$^{-1}$, also ascribed to a pure amorphous Zr [58,69,78]. The values of $\chi_{exp}$ of (TiZrNbCu)$_{1-x}$Ni$_x$ and (TiZrNbNi)$_{1-x}$Cu$_x$ MGs also extrapolated close to this value for $x$=0 [17, 18, 24, 27]. In addition to comparable initial values, the data for both systems in Fig. 9 show also similar variations with $x$: after a small initial decrease on increasing $x$ their $\chi_{exp}$ rapidly increase at some elevated $x$. In our alloys, this occurs for $x$>0.25, where, also, the variations of the thermal parameters (Fig. 4), average atomic volume and average local packing fraction (Fig. 7) showed some change, whereas in Zr-Co ones this happens for $x>0.4$ where their $N(E_F)$ also starts to increase on increasing $x$ [80] and is close to concentration at which the band crossing occurs in these MGs [48,49]. Moreover, the variations of $\chi_{exp}$ and $N(E_F)$ with $x$ in $Zr_{1-x}Co_x$ MGs are qualitatively the same [80]. Despite the apparent similarity between the variations of $\chi_{exp}$ and $N(E_F)$ observed in several TE-TL MGs [17,18,21,22,24,27,46,58-61,69,77,80] the magnetic susceptibility of transition metals and alloys is quite complex and consists of three main contributions of which the only one is related to $N(E_F)$ [46,69,77]:

$$\chi_{exp}=\chi_p+\chi_{orb}+\chi_{dia} \qquad (6)$$

where $\chi_p$ is the Pauli paramagnetic contribution, of d-electrons, and $\chi_{orb}$ and $\chi_{dia}$ are the orbital paramagnetic and diamagnetic contributions, respectively. $\chi_{orb}$ and $\chi_{dia}$ of an alloy are obtained by adding corresponding contributions from the alloying constituents [77,81,82]. $\chi_p$ is enhanced over the free-electron value, $\chi_p^0$ (proportional to $N(E_F)$) by the exchange interaction: $\chi_p=S\chi_p^0$, where $S$ is the Stoner enhancement which also depends on $N(E_F)$ [46,69]. Since the orbital contribution (dominated with large contributions of TEs [46,58,69,77]) varies linearly with TL content and diamagnetic contribution is generally smaller and varies only a little with composition it is the $\chi_p$ term that determines the dependence of $\chi_{exp}$ on concentration. Further, in some TE-TL MGs, such as those containing TL=Cu or Ni S is nearly constant over a broad concentration range with result that $\chi_p$ is proportional to $N(E_F)$ [17,18,27,58,69,78] like that in normal paramagnetic metals and alloys [71]. However, nearly constant $\chi_{exp}$ of $Zr_{1-x}Co_x$ alloys for $x\leq0.4$, despite of decreasing $N(E_F)$ over the same concentration range was ascribed to an increase of S with x [58,69] due to the vicinity to the threshold for long-range ferromagnetism in this system around $x=0.6$ [79,80].

Since we have neither the values of $N(E_F)$ for all $(TiZrNbCu)_{1-x}Co_x$ alloys [51], nor the corresponding superconducting transition temperatures (e.g.,[29]) we cannot calculate $\chi_p^0$, thus the variation of $S$ with $x$ in our alloys. Therefore, we cannot properly analyze the origin of the enhancement of $\chi_{exp}$ for $x>0.25$ in Fig. 9. Such an early increase of $\chi_{exp}$ is quite surprising considering a rather large concentration for band crossing estimated for our system, around 0.68. Further, a considerable decrease of $\chi_{exp}$ in going from $x=0$ to $x=0.25$ and a relatively small increase of $N(E_F)$ from $x=0.2$ to $x=0.43$ seem to indicate that $N(E_F)$ continues to decrease with increasing $x$ possibly beyond $x=0.25$. Accordingly, a slow increase of $S$ with $x$, similar to that calculated in $Zr_{1-x}Co_x$ MGs [58,69] would not be sufficient to explain a sudden increase of $\chi_{exp}$ at $x=0.35$ in Fig. 9. However, our partially crystalline alloy with x=0.5 showed ferromagnetic hysteresis loops at low temperatures, T=2K and 30K. Thus, the threshold for the onset of ferromagnetism in our alloys seems lower than that in Zr-Co MGs [79,80]. Accordingly, the Stoner enhancement S should increase more rapidly in our alloys than that in Zr-Co MGs which can explain the observed enhancement of $\chi_{exp}$ at relatively low Co content (Fig. 9).

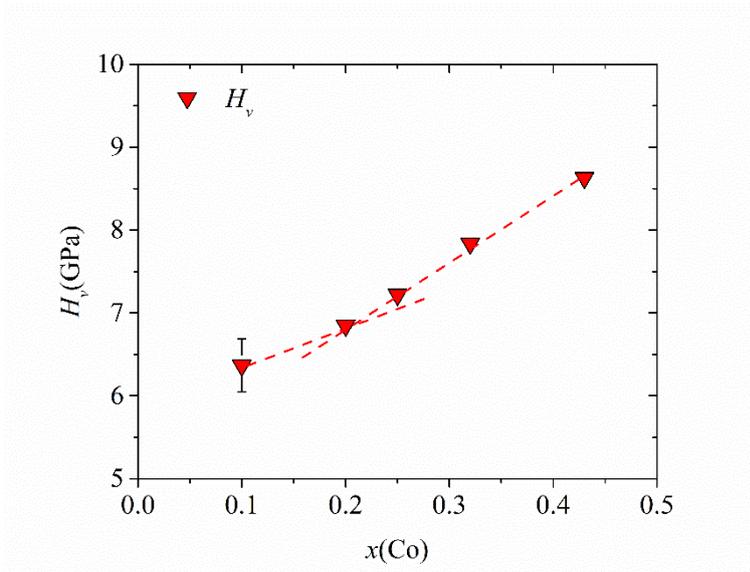

Fig. 10. Microhardness $H_v$ of as-cast $(TiZrNbCu)_{1-x}Co_x$ alloys vs. $x$.

Metallic glasses show some outstanding mechanical properties, such as very large strength (approaching the theoretical strength) as well as very high hardness, fracture toughness and wear resistance which make them interesting for diverse practical applications [83]. The very large hardness of MGs is associated with the absence of extended crystal defects (hence the absence of crystal-slip mechanisms) which results in the elastic-plastic type of deformation, thus the yield stress practically the same as the tensile limit. Accordingly, their hardness is proportional to both, the yield stress and the Young's modulus $E$ [84] and since MGs are macroscopically homogenous and isotropic it is also proportional to their shear and bulk modules, $G$ and $B$, respectively. The absence of crystal-slip mechanisms makes all these properties directly related to the strength of interatomic bonding [17,18,27,46,58-62]. Since this strength also affects the average atomic volume and the thermal parameters $T_g$ and $T_x$, all these parameters are also correlated with the mechanical properties of MGs [18,27,58-62,85].

Variation of the room temperature microhardness, $H_V$ of our as-cast $(TiZrNbCu)_{1-x}Co_x$ with Co content is shown in Fig. 10. To our knowledge, these are the first results for the variation of $H_V$ on crossing from HEA to conventional alloy concentration range. Strong increase of $H_V$ with x is consistent with the corresponding increase of the interatomic bonding showing up in an increase of $\Delta H_{mix}$ with $x$ in Fig. 1, as well as that of thermal parameters $T_g$, $T_x$ and $T_m$ in Fig. 4. Further, the $H_V(x)$ variation is apparently correlated with the variations of the average atomic volume (Fig. 7) and the mass density. Indeed, it seems plausible that in alloys with a similar atomic structure (amorphous) hardness increases with density. Moreover, a subtle change in the local atomic arrangements which affects the variations of the average atomic volumes, the atomic packing fractions and the average coordination number of the first coordination shell, $N$ [50] around $x$=0.25 (Fig. 7) seem to affect the variation of $H_V$, which increases more rapidly with x for x>0.2. Although an increase in the atomic packing fraction and in $N$ should increase the hardness rather large experimental error inherent to measurement of $H_V$ (Fig. 10) makes it possible to draw also single straight line through data in Fig. 10.

Since there are no results for $N(E_F)$ of most of our alloys [51] the discussion of the variation of $H_V$ in terms of the corresponding change in ES would be at present premature. However, due to the tendency of band crossing in our alloys on increasing $x$ (which shows up in the evolution of the photoemission spectra in Fig. 8 and the increase of $N(E_F)$ for $x$=0.43 [51]) we expect that at elevated $x$ the mechanical properties of our alloys will approach those of pure cobalt.

In the absence of the results for the mechanical properties of TE-Co MGs we will compare our results with those for $H_V$ and $E$ of Ti,Zr,Hf-Ni,Cu MGs [46,58-61,84,86] and with results for $E$ of (TiZrNbCu)$_{1-x}$Ni$_x$ alloys [17,24,27]. Both, the magnitudes of $H_V$ and the average slope of their increase with $x$ in Fig. 10 are larger than those observed in binary TE-Ni,Cu MGs. In particular, in equiatomic Ti-Ni,Cu and Zr-Ni,Cu MGs $H_V$ is around 7 GPa and 5 GPa respectively, whereas in our alloys $H_V$ extrapolates to about 9 GPa for $x$=0.5 (Fig. 10). Similarly, $H_V$ of our alloys extrapolates to 5.8 GPa for $x$=0, whereas those for Ti-Ni,Cu and Zr-Ni,Cu extrapolate to those of amorphous Ti and Zr around 5 and 3.5 GPa, respectively [46,58-60]. Linear extrapolation of our results to $x$=1 yields $H_V$ around 12 GPa for a pure, amorphous Co which is, as expected much larger than that of the crystalline phase of Co. To compare our results with those for $E$ of binary and quinary TE-TL MGs we will assume that the universal relationship between $E$ and $H_V$ of metal-metal type (M-M) binary and ternary MGs [84], $E$=15 $H_V$ is applicable to our alloys. According to this expression, $E$ of our alloys extrapolates to about 88 GPa for $x$=0, which is close to that obtained by extrapolation of the experimental results for $E$ of (TiZrNbCu)$_{1-x}$Ni$_x$ MGs to $x$=0 [17,24,27]. The magnitude of $E$ obtained by conversion of the linear extrapolation of $H_V$ to $x$=0.5 in Fig. 10 is around 135 GPa, which is considerably larger than E=115 GPa measured for (TiZrNbCu)$_{0.5}$Ni$_{0.5}$ MG [27], but is very close to that of polycrystalline ZrCo compound, E=132 GPa [87]. Similarly, the extrapolated value of $H_V$ to x=1 corresponds approximately to E=190 GPa for amorphous Co, which is as expected somewhat lower, but close to that of a pure Co. Thus, the empirical proportionality between $E$ and $H_V$ in M-M MGs seems to be a reasonable approximation also in our alloys. For this type of MGs, one can also estimate the yield stress from $H_V$ [84] and, providing that one knows the Poisson's ratio, calculate the values of $G$ and $B$ from that of $E$. Since the vibrational properties, in particular the Debye temperatures, of MGs (e.g., [17,24,27]) are also related to the strength of interatomic bonding, the observed increase of the Debye temperature in our alloys going from $x$=0.2 to 0.43 [51] also supports the increase in the strength of interatomic bonding with increasing Co content. Finally, we note that the Co-based metal-metalloid type of MGs with a high content of boron have the largest strengths and $H_V$s (≤16.1 GPa) of all known MGs [88]. However, in this type of MGs the mechanism of interatomic bonding is different from that in metal-metal ones [89].

4. Conclusion

A comprehensive study of the transition from the regime of high-entropy (HEA) to that of conventional alloys based on Co has been performed in a new amorphous $(TiZrNbCu)_{1-x}Co_x$ alloy system. Careful analysis of the nonequilibrium phase diagram, X-ray diffraction patterns and magnetic susceptibility shows that this system is unique in the sense that all parameters change their variation with composition around $x=0.25$ ($VEC=6.7$), thus deep within the HEA (bcc) regime.

Indeed, in all other quinary alloy systems which have been studied in some detail over an extended concentration range, such as Ti-Zr-Nb-Cu-Ni metallic glasses (MG) [17,18,27,29] and fcc crystalline Cr-Mn-Fe-Co-Ni alloys [35-38] there was either no change in the variation of properties within the explored concentration range ( fcc $(CrMnCoNi)_{1-x}Fe_x$ and $(CrMnFeNi)_{1-x}Co_x$ alloys [38] and $(TiZrNbNi)_{1-x}Cu_x$ MGs [18]) , or the change in variation occurred within the regime of conventional alloys ( fcc $(CrMnFeCo)_{1-x}Ni_x$ alloys [35,38] and $(TiZrNbCu)_{1-x}Ni_x$ MGs [17,27,29]).

In present MGs thermal parameters (Fig. 4), such as the glass transition, crystallization and liquidus temperatures, change the slope of their increase on increasing Co content around $x=0.25$. Careful analysis of thermal parameters provided a new possible explanation of the well-known discrepancy between high thermal stability and rather poor glass-forming ability of HE-MGs [24,56] in terms of their complex crystallization patterns. This discrepancy can simply be due to fact that different crystallization events, thus crystalline phases, determine the thermal stability of glass and that of melt. The average atomic volumes, obtained from a detailed analysis of X-ray diffraction patterns, show a deviation from the Vegard´s law for $x≤0.25$, while the average local atomic packing fraction suddenly increases above this composition, Fig. 7. (These findings have very recently been corroborated by the results of the study of radial distribution functions of the same alloys, performed using the high energy X-ray diffraction, which showed a slower change of the average interatomic distance with Co content for $x≤0.25$, and an increase in the average number of neighbours within the first coordination shell for $x≥0.25$ [50].) Such a change in local atomic arrangements in the amorphous state and within the HEA concentration region is quite unusual.

The magnetic susceptibility of the same alloys (Fig. 9) passes through the minimum value at $x=0.25$ and rapidly increases with increasing Co content above that concentration. The initial decrease and then the increase of magnetic susceptibility with Co content is qualitatively consistent with the split-band shape of the valence band (observed using ultraviolet photoemission spectroscopy, UPS) and the shift of 3d states of Co towards the Fermi level on increasing $x$ in these alloys, Fig. 8. The preliminary results of the specific heat measurements [51] indicate an increase of the density of states at the Fermi level for $x=0.43$, thus are consistent with both UPS and magnetic susceptibility results.

The first study of the variation of microhardness across the transition from HEA to conventional Co-based concentration range revealed very large hardness that increases rapidly on increasing Co content. Present results also indicate that the correlation between the microhardness and elastic modules, well established in conventional, binary MGs of metal-metal type [46,60,84], may apply to ours and other compositional complex metal-metal type amorphous alloys.

The observed possibility to tune the local atomic arrangements (within the amorphous phase) and physical properties by changing the composition and/or selected principal alloying constituent (e.g., Cu for Ni or Co) may be useful for the fabrication of amorphous alloys and coatings with predetermined properties. Moreover, recent results for Cantor type of alloys [35-38] indicate that a similar approach is applicable to single-phase crystalline alloys, too. This emphasizes the importance of studying the transition from HEA to conventional alloys with the same chemical make-up.

Acknowledgement

We acknowledge the support of project CeNIKS co-financed by the Croatian Government and the European Union through the European Regional Development Fund - Competitiveness and Cohesion Operational Programme (Grant no. KK.01.1.1.02.0013). I. A. Figueroa acknowledges the financial support of UNAM-DGAPA-PAPIIT.